\documentclass[10pt,twocolumn,twoside]{IEEEtran}

\usepackage[utf8]{inputenc} 
\usepackage[T1]{fontenc}
\usepackage{url}
\usepackage{ifthen}
\usepackage{cite}
\usepackage[cmex10]{amsmath} 
\usepackage{amsmath,amsthm,amssymb}
\usepackage{tikz}
\usepackage{bm}
\usepackage{mathdots}
\usepackage{pgfplots}
\pgfplotsset{compat=1.15}
\usepackage{mathrsfs}
\usetikzlibrary{arrows}
\usepackage{yhmath}
\usepackage{cancel}
\usepackage{color}
\usepackage{siunitx}
\usepackage{array}
\usepackage{multirow}
\usepackage{hyperref}
\usepackage{mathtools}
\usepackage{amssymb}
\usepackage{caption}
\usepackage{subcaption}
\usepackage{gensymb}
\usepackage{tabularx}
\usepackage{mathtools}

\DeclarePairedDelimiter\floor{\lfloor}{\rfloor}
\usepackage{setspace}

\theoremstyle{definition}
\newtheorem{definition}{Definition}
\newtheorem{remark}{Remark}
\newtheorem{example}{Example}
\newtheorem{theorem}{Theorem}

\newtheorem{lemma}{Lemma}

\title{The Discrepancy Attack on Polyshard-ed Blockchains} 
\author{%
  Nastaran Abadi Khooshemehr, Mohammad Ali Maddah-Ali \\
  Sharif University of Technology \\
  Email: abadi.nastaran@ee.sharif.ir,  maddah\_ali@sharif.edu
}
\begin{document}



\maketitle

\begin{abstract}
Sharding, i.e. splitting the miners or validators to form and run several subchains in parallel, is known as one of the main solutions to the scalability problems of blockchains. The drawback is that as the number of miners expanding each subchain becomes small, it becomes vulnerable to security attacks. To solve this problem, a framework, named as \textit{Ployshard}, has been proposed in which each validator verifies a coded combination of the blocks introduced by different subchains, thus helping to protect the security of all subchains.  In this paper, we introduce an attack on Ployshard, called \textit{the discrepancy} attack, which is the result of malicious nodes controlling a few subchains and dispersing different blocks to different nodes. We show that this attack undermines the security of Polyshard and is undetectable in its current setting.
 
\end{abstract}


\section{Introduction}
Blockchain is known as a disruptive technology that would eliminate the necessity of trusting centralized entities, running, and sometimes abusing,  the financial and information networks. For this promise of blockchain systems to be fulfilled, the fundamental problems regarding the underlying technology should be addressed. In that sense, there has been a remarkable effort in the community to solve the \textit{blockchain trilemma}.  This trilemma highlights the challenge of simultaneously  achieving three main requirements  of a blockchain network: (1) Scalability, i.e., the throughput of the blockchain with $N$ nodes should scale with $N$, (2) Security, i.e. the blockchain network should function properly even if a constant fraction of the nodes is controlled by adversaries, (3) Decentralization, i.e. the storage, communication, and computation resources needed by each node  remain constant, as $N$ increases. 
For a survey on approaches to this problem, refer to \cite{zhou2020solutions}.

One of the promising approaches to blockchain scalability problem is \textit{sharding}. In sharding, the system exploits the gain of parallel processing by splitting the nodes into several shards, where each shard forms and maintain a subchain, in parallel with other shards. 
The drawback is that as the number of nodes maintaining a subchain becomes small, it becomes vulnerable to security attacks. 
Recently, in~\cite{li2020polyshard}, a scheme, named \emph{Polyshard}, has been proposed that claims to achieve  linear scaling in  throughput and security threshold. In addition, the storage and computation resources per node remain constant. This would be a remarkable step toward outlining a solution to the blockchain trilemma.  The main idea of Polyshard is incorporating coded computing techniques in the context of blockchain.

Coded computing is based on running the desirable computation tasks on some linear combinations of the inputs,  rather than each input individually~\cite{speed, yu2017polynomial, opt-recovery, yu2020straggler, codedsketch, yu2019lagrange, Hanzaleh}.  This would help the system to detect and correct the results of adversarial nodes,  ignore the results of straggles, and even protect the input data against curious nodes.   In particular, in~\cite{yu2019lagrange},   Lagrange coded computing (LCC) has been introduced to compute a  general polynomial function of several inputs on a cluster of servers, where some of them are adversaries. In Polyshard, Lagrange coded computing is used to run the function, verifying the validity of the blocks, on some linear combinations of the blocks produced by the subchains. It is claimed that this would  entangle the security of all shards together and improve the security of the blockchain, without increasing the computation and storage cost at each node. Blockchain systems can benefit from coding in other ways as well. For example, \cite{kadhe2019sef}, \cite{wu2020distributed}, \cite{perard2018erasure} use coding to reduce storage in nodes, \cite{pal2020fountain} uses coding for easing the bootstrap process of new nodes, and \cite{al2018fraud,yu2020coded} use coding to tackle the availability problem in blockchain.

In this paper, we introduce a fundamental attack on Polyshard that undermines its security. The attack is based on an adversarial behavior in the system that is unobserved in \cite{li2020polyshard}. In essence, the adversarial nodes can take the control of a few shards and 
transmit inconsistent blocks to different nodes. The heavy load of communication does not allow nodes to resolve the inconsistencies. 
These inconsistent versions make the set of equations used for decoding inconsistent. We prove that this inconsistency cannot be resolved by linear decoding, unless  the number of nodes $N$ grows at least linear with $K^{\frac{d}{2}}$, where $K$ is the number of shards and $d$ is the degree of the verification function of the blockchain system. This prevents the system from tolerating $O(N)$ adversaries. The general discrepancy attack that results from some malicious nodes distributing inconsistent data, can happen in many systems and is not limited to blockchains. We have studied the fundamental problem of distributed encoding in \cite{abadi2020fundamental}.

In the following, first, we summarize Polyshard in Section \ref{summary of polyshard}. Then, we explain the attack in Section \ref{attack section}, and analyze it in Section \ref{attack analysis section}.

\section{Summary of Polyshard \cite{li2020polyshard}}\label{summary of polyshard}
In this section, we review Polyshard \cite{li2020polyshard}, and adopt the notation in \cite{li2020polyshard}. Each shard $k\in[K]$ has a subchain, denoted by $Y^{t-1}_k=(Y_k(1),\dots,Y_k(t-1))$ right before epoch $t\in\mathbb{N}$, where $Y_k(t)\in\mathbb{U}$ denotes the block accepted to shard $k$ at epoch $t$, and $\mathbb{U}$ is a vector space over the finite field $\mathbb{F}$. All computations are in $\mathbb{F}$. It is assumed that in each epoch, a new block $X_k(t)\in\mathbb{U}$ is \textit{proposed} to each shard $k$. In the next section, we will question this assumption, but for now, let us accept it. The new blocks should be verified by a verification function $f^t:\mathbb{U}^t\rightarrow\mathbb{V}$ with $X_k(t)$ and $Y_k^{t-1}$ as inputs, where $\mathbb{V}$ is a vector space over $\mathbb{F}$. This function depends on the consensus algorithm of the blockchain, but can be expressed as a multivariate polynomial in general. 

In order to verify $X_k(t)$, $k\in[K]$, we need $h_k^t=f^t(X_k(t),Y_k^{t-1})$, so that we can compute $e_k^t=\mathbf{1}(h_k^t\in\mathcal{W})$, and then the verified block $Y_k(t)=e_k^tX_k(t)$, where $\mathcal{W}\subseteq \mathbb{V}$ denotes the set of function outputs that affirm $X_k(t)$, and $\mathbf{1}$ is the indicator function. In other words, we need to calculate $f^t(X_1(t),Y_1^{t-1}),\dots,f^t(X_K(t),Y_K^{t-1})$ to verify the incoming blocks.

In polyshard, each shard $k\in[K]$ and each node $n\in[N]$ are associated with distinct constant values $\omega_k\in\mathbb{F}$, and $\alpha_n\in\mathbb{F}$, respectively. There are two global Lagrange polynomials for each epoch $m\in\mathbb{N}$,
\begin{align}
    p_m(z) = \sum_{k=1}^K Y_k(m)\prod_{j\neq k}\frac{z-\omega_j}{\omega_k-\omega_j}, \\
    q_m(z) = \sum_{k=1}^K X_k(m)\prod_{j\neq k}\frac{z-\omega_j}{\omega_k-\omega_j}.
\end{align}
These polynomials are such that $p_m(\omega_k)=Y_k(m)$ and $q_m(\omega_k)=X_k(m)$, for all $k\in[K]$. Thus, $f^t(X_k(t),Y_k^{t-1})$, $k\in[K]$ are equivalent to $f^t(q_t(\omega_k),p_1(\omega_k),\dots,p_{t-1}(\omega_k)), k\in[K]$. 

At the beginning of epoch $t$, a node $n\in[N]$ has the coded chain $\Tilde{Y}_n^{t-1}=(\Tilde{Y}_n(1),\dots,\Tilde{Y}_n(t-1))$ in its storage, where 
\begin{align*}
    \Tilde{Y}_i(m)=p_m(\alpha_n)=\sum_{k=1}^K Y_k(m)\prod_{j\neq k}\frac{\alpha_n-\omega_j}{\omega_k-\omega_j},~ m\in[t-1].
\end{align*}
This coded chain is the outcome of executing Polyshard in the previous epochs. All nodes in the network receive the newly proposed blocks $X_1(t),\dots,X_K(t)$. Each node $n\in[N]$ calculates the coded block $\Tilde{X}_n(t)=q_t(\alpha_n)$, and then verifies the coded block against the stored coded chain, i.e. calculates $f^t(\Tilde{X}_n(t),\Tilde{Y}_n^{t-1})\coloneqq g_n^t$, and broadcasts the result in the network. After this step, all nodes have $g_1^t,\dots,g_N^t$, though the values from the adversarial nodes may be arbitrary or empty. Since
\begin{align*}
    g_j^t=f^t(q_t(\alpha_j),p_1(\alpha_j),\dots,p_{t-1}(\alpha_j)),\quad j\in[N],
\end{align*}
and each node $n$ needs $f^t(q_t(\omega_k),p_1(\omega_k),\dots,p_{t-1}(\omega_k))$ for all $k\in[K]$, node $n$ should determine the polynomial 
\begin{align*}
    f^t(q_t(z),p_1(z),\dots,p_{t-1}(z))
\end{align*}
using $g_1^t,\dots,g_N^t$, and then evaluate it at $\omega_1,\dots,\omega_K$ to find 
$f^t(X_1(t),Y_1^{t-1}),\dots,f^t(X_K(t),Y_K^{t-1})$.

The degree of $f^t(q_t(z),p_1(z),\dots,p_{t-1}(z))$ is $d(K-1)$, so there are $d(K-1)+1$ unknowns to be found. The Reed-Solomon decoding allows a maximum of $\floor*{\frac{N-(d(K-1)-1)}{2}}$ incorrect values among $g_1^t,\dots,g_N^t$.
If $\mu$ fraction of the nodes are controlled by the adversary (i.e. $\beta=\mu N$), it suffices to have $\mu N\le \floor*{\frac{N-(d(K-1)-1)}{2}}$, which means $K=O(N)$ is feasible.

Polyshard is indeed based on Lagrange coded computing in \cite{yu2019lagrange}. In LCC, there are $N$ workers, and a master that wants $f(X_1),\dots,f(X_K)$. Worker $i\in[N]$ is provided with $\Tilde{X}_i$, which is an encoded version of $X_1,\dots,X_K$. The encoding is done with a Lagrange polynomial, exactly similar to Polyshard. Workers apply $f$ on their encoded data and send them to the master. The master can recover $f(X_1),\dots,f(X_K)$ using Reed-Solomon decoding.

\section{The discrepancy Attack}\label{attack section}
In this section, we explain the discrepancy attack on Polyshard and show that it undermines both the security and scalability of Polyshard. 

The attack stems from the assumption that all nodes in the network receive the same new blocks $X_1(t),\dots,X_K(t)$ at the beginning of each epoch $t\in\mathbb{N}$. As mentioned in the previous section, Polyshard assumes that new blocks are somehow \textit{proposed} to shards. But in fact, the new block for a shard is produced by nodes in that shard, and not through an external process. 
An adaptive adversary can decide to control an arbitrary subset of the nodes up to a certain size.
When a certain number of nodes in shard $k\in[K]$ are under the control of the adversary, they can produce more than one new block $X_k(t)$, say $X_k^{(1)}(t),\dots,X_k^{(v)}(t)$, for some $v\in\mathbb{N}$, and send them to different nodes in the network. Note that the adversaries do not \textit{broadcast} the produced blocks, instead, they use separate links to deliver the new blocks to different nodes so that nodes would not know what other nodes have received and remains unaware of this attack. By doing so, the adversary can cause a sort of \textit{discrepancy} in the network while remaining undetected by the nodes.

In polyshard, the only considered adversarial behavior is broadcasting incorrect $g_n^t$, which is subdued in the process of decoding. But an adversary need not wait till that step and can violate the protocol in the very first step by sending inconsistent data to different nodes. In the following, we investigate the effects of the discrepancy attack.

Suppose that only one shard is controlled by the adversary, and the other $K-1$ shards are honest (contain only honest nodes). We show that even one adversarial shard can have detrimental effects. Let the adversarial shard be the first shard, producing $v\in\mathbb{N}$ blocks $X_1^{(1)}(t),\dots,X_1^{(v)}(t)$ for epoch $t$. Some or all of these blocks may contradict the history of the first shard. We assume that the adversary attacks at epoch $t$ for the first time.
Honest shards produce and broadcast $X_2(t),\dots,X_K(t)$. We define the polynomials
\begin{align}
    q^{(i)}_m(z) &= X_1^{(i)}(m)\prod_{j\neq 1}\frac{z-\omega_j}{\omega_1-\omega_j}+ X_2(m)\prod_{j\neq 2}\frac{z-\omega_j}{\omega_2-\omega_j} \nonumber \\
    & +\dots +X_K(m)\prod_{j\neq K}\frac{z-\omega_j}{\omega_K-\omega_j}, \quad i\in[v],
\end{align}
for epoch $m\in\mathbb{N}$.
Suppose that node $n\in[N]$ receives $X_1^{(v_n)}$, where $v_n\in[v]$. Following Polyshard, node $n$ first calculates $\Tilde{X}_n(t)=q^{(v_n)}_t(\alpha_n)$ (not knowing $v_n$, and whether any attack is going on), then calculates $g_n^t = f^t(\Tilde{X}_n(t),\Tilde{Y}_n^{t-1})=f^t(q_t^{(v_n)}(\alpha_n),p_1(\alpha_n),\dots,p_{t-1}(\alpha_n))$, and then broadcasts $g_n^t$. The adversarial nodes may broadcast arbitrary values or may broadcast nothing. 

After the broadcast, all nodes have $g_1^t,\dots,g_N^t$, but unlike the previous section, they are not points on a single polynomial. In other words, there are $v$ different polynomials,
\begin{align*}
    f^t(q_t^{(1)}(z),p_1(z),&\dots,p_{t-1}(z)),\\
    &\vdots \\
    f^t(q_t^{(v)}(z),p_1(z),&\dots,p_{t-1}(z))
\end{align*}
and we have $N$ evaluations of these polynomials at $\alpha_1,\dots,\alpha_N$. Therefore, the Reed-Solomon decoding in Polyshard fails and cannot find any of the above polynomials. 

To realize the severity of the discrepancy attack, let us consider a simple solution that aims to make nodes aware of the attack so that honest nodes can set the blocks of the adversarial nodes aside and continue with their own blocks. Assume that all nodes broadcast their $K$ received blocks so that each node sees what other nodes have received and gets notified of any inconsistency. This entails an overall communication load of $O(N^2K)$, but $K=O(N)$, and the communication load of $O(N^3)$ is unbearable\footnote{It is worth noting that the communication load of Polyshard for each node is $O(N)$, which troubles the decentralization property.}.

The discrepancy attack can harm the underlying LCC engine of Polyshard if distributed encoding is deployed in it. In LCC, the encoded data in workers is assumed to be the outcome of a centralized encoding entity that receives the raw data, encodes them, and distributes the encoded data among the workers. If encoding is done in a distributed manner, e.g. workers receive the raw data and perform the encoding themselves as in Polyshard, LCC becomes prone to the discrepancy attack. We have studied the problem of distributed encoding in \cite{abadi2020fundamental}.

\section{Analysis of the Discrepancy Attack}\label{attack analysis section}
In this section, we study the more general problem of Lagrange coded computing with distributed encoding, with Polyshard as a special case. 

There are $N$ nodes, $\beta$ of which are adversarial. We denote the set of all nodes with $\mathcal{N}=\{1,\dots,N\}$, and the set of adversarial and honest nodes with $\mathcal{A}$ and $\mathcal{H}$ respectively, where $\mathcal{N}=\mathcal{A}\cup\mathcal{H}$. The first $K$ nodes in $\mathcal{N}$, i.e. the set $\mathcal{K}=\{1,\dots,K\}$ are message producers, where $\mathcal{K}=\mathcal{A}_{\mathcal{K}}\cup\mathcal{H}_{\mathcal{K}}$, and $\mathcal{A}_{\mathcal{K}}=\mathcal{A}\cap\mathcal{K}, \mathcal{H}_{\mathcal{K}}=\mathcal{H}\cap\mathcal{K}$. Let the number of adversarial message producers be $\beta'$, i.e. $|\mathcal{A}_{\mathcal{K}}|=\beta'$.

Node $k\in\mathcal{H}_{\mathcal{K}}$ sends $X_k\in\mathbb{U}$ to all the other nodes, while node $k\in\mathcal{A}_{\mathcal{K}}$ sends $X_k^{(\nu_{k,n})}\in\mathbb{U}$, $\nu_{k,n}\in[v]$ to node $n\neq k$, where $\mathbb{U}$ is a vector space over $\mathbb{F}$. Each adversarial node can inject at most $v\in\mathbb{N}$ different messages in the network, i.e. $|\{X_k^{(\nu_{k,n})}, n\in\mathcal{N}\}|\le v$ for all $k\in\mathcal{A}_{\mathcal{K}}$. Indeed, we consider a (constant) maximum on the number of different messages that an adversarial node can inject into the system\footnote{This assumption is necessary, because too many versions from even one adversary can destroy the system. It is also in accordance with practical system.}. Honest nodes do not know which nodes are adversarial, and how adversarial nodes send their messages.
Each node $n\in\mathcal{N}$ does the following in order:
\begin{enumerate}
    \item calculates a coded version of the received data using Lagrange \textit{encoder polynomial},
    \begin{align}\label{q^V definition}
    q^{(\mathbf{v})}(z) \coloneqq\sum_{k\in\mathcal{A}_{\mathcal{K}}} X_k^{(\mathbf{v}[k])}\prod_{j\neq k}\frac{z-\omega_j}{\omega_i-\omega_j}+ \sum_{k\in\mathcal{H}_{\mathcal{K}}} X_k\prod_{j\neq k}\frac{z-\omega_j}{\omega_i-\omega_j},
\end{align}
where $\mathbf{v}=(\mathbf{v}[k])_{ k\in\mathcal{A}_{\mathcal{K}}}\in[v]^{\beta'}$ is a $\beta'$-tuple of elements in $[v]$, denoting the versions of the received adversarial messages, and $\omega_k\in\mathbb{F},k\in[K]$ are distinct constants assigned to message producer nodes. Then, node $n$ evaluates $q^{(\mathbf{v})}(z)$ at $\alpha_n\in\mathbb{F}$, $\Tilde{X}_n=q^{(\mathbf{v})}(\alpha_n)$.
    \item calculates $y_n=f(\Tilde{X}_n)$, where $f:\mathbb{U}\rightarrow \mathbb{V}$ is a polynomial of degree $d\in\mathbb{N}$, and $\mathbb{V}$ is a vector space over $\mathbb{F}$.
    \item broadcasts $y_n$.
\end{enumerate}
The honest nodes should be able to recover $f(X_k), k\in\mathcal{H}_{\mathcal{K}}$, after they receive enough number of $y$'s. We define the recovery threshold $N^*$, such that for \textit{any} set $\mathcal{N}^*\subseteq\mathcal{N}$, $|\mathcal{N}^*|=N^*$, $f(X_k), k\in\mathcal{H}_{\mathcal{K}}$ can be decoded from $\{y_n, n\in\mathcal{N}^*\}$.

In order to state our result, we need the following definitions.
\begin{definition}[coefficient vector]
Let the Lagrange encoder polynomial be 
\begin{align}\label{encoding polynomial q}
    q^{(\mathbf{v})}(z) = \sum_{i=0}^{K-1}L_i(X_r^{(\mathbf{v}[r])},X_s,~ r\in \mathcal{A}_{\mathcal{K}},s\in \mathcal{H}_{\mathcal{K}})z^i,
\end{align}
where $L_0,\dots,L_{K-1}:\mathbb{F}^{K}\rightarrow\mathbb{F}$ are linear maps, and 
\begin{align}\label{foq expansion}
    f(q^{(\mathbf{v})}(z)) = \sum_{i=0}^{d(K-1)}u_i(X_r^{(\mathbf{v}[r])},X_s, r\in \mathcal{A}_{\mathcal{K}},s\in \mathcal{H}_{\mathcal{K}})z^i,
\end{align}
where $u_0,\dots,u_{d(K-1)}:\mathbb{F}^{K}\rightarrow\mathbb{F}$ are polynomials of degree $d$ of $X_r^{(\mathbf{v}[r])}, r\in\mathcal{A}_{\mathcal{K}}$ and $X_s,s\in\mathcal{H}_{\mathcal{K}}$.
The \textit{coefficient vector} of this encoding, denoted by $\mathbf{U}_{\textrm{Lagrange}}$ is defined as \begin{align}\label{U definition}
    \mathbf{U}_{\textrm{Lagrange}} \coloneqq \left[\begin{array}{c}
    u_{d(K-1)}(X_r^{(\mathbf{v}_1[r])},X_s,~ r\in \mathcal{A}_{\mathcal{K}},s\in \mathcal{H}_{\mathcal{K}}) \\
    \vdots \\
    u_{0}(X_r^{(\mathbf{v}_1[r])},X_s,~ r\in \mathcal{A}_{\mathcal{K}},s\in \mathcal{H}_{\mathcal{K}}) \\ \hline
    \vdots \\ \hline
    u_{d(K-1)}(X_r^{(\mathbf{v}_{v^{\beta'}}[r])},X_s,~ r\in \mathcal{A}_{\mathcal{K}},s\in \mathcal{H}_{\mathcal{K}}) \\
    \vdots \\
    u_{0}(X_r^{(\mathbf{v}_{v^{\beta'}}[r])},X_s,~ r\in \mathcal{A}_{\mathcal{K}},s\in \mathcal{H}_{\mathcal{K}})
    \end{array}\right],
\end{align}
where $\mathbf{v}_1,\dots,\mathbf{v}_{v^{\beta'}}$ are all the $v^{\beta'}$ $\beta'$-tuples of the elements of $[v]$.
\end{definition}
The length of $\mathbf{U}$ is $v^{\beta'}(d(K-1)+1)$.
\begin{example}\label{example 1}
Assume $K=N=3, \beta=2, v=2, \mathcal{A}=\{1,2\},\mathcal{H}=\{3\},(\omega_1,\omega_2,\omega_3)=(1,2,3)$, and $f(x)=x^2$.
\begin{table*}
    \centering
    \begin{align}
    q^{([i_1,i_2])}(z)&=\frac{(z-2)(z-3)}{2}X^{(i_1)}_1 + \frac{(z-1)(z-3)}{-1}X_2^{(i_2)} +\frac{(z-1)(z-2)}{2}X_3 \nonumber \\
    & = z^2(\frac{X_1^{(i_1)}}{2}-X_2^{(i_2)}+\frac{X_3}{2})+ z(-\frac{5X_1^{(i_1)}}{2} +2X_2^{(i_2)} -\frac{3X_3}{2})+(3X_1^{(i_1)}-3X_2^{(i_2)}+X_3), \quad i_1,i_2\in\{1,2\},
\end{align}
    \begin{align}\label{f o u in example 1}
f(u^{([i_1,i_2])}(z)) &=\bigg(\frac{(X_1^{(i_1)})^2}{4}-X_1^{(i_1)}X_2^{(i_2)}+\frac{X_1^{(i_1)}X_3}{2}+(X_2^{(i_2)})^2  -X_2^{(i_2)}X_3+\frac{X_3^2}{4}\bigg)z^4 + \bigg(\frac{-5(X_1^{(i_1)})^2}{2}+9X_1^{(i_1)}X_2^{(i_2)} \nonumber \\
& -4X_1^{(i_1)}X_3 -8(X_2^{(i_2)})^2+7X_2^{(i_2)}X_3-\frac{3X_3^2}{2}\bigg)z^3 + \bigg(\frac{37(X_1^{(i_1)})^2}{4}-29X_1^{(i_1)}X_2^{(i_2)}  +\frac{23X_1^{(i_1)}X_3}{2}+22(X_2^{(i_2)})^2 \nonumber \\
& -17X_2^{(i_2)}X_3 +\frac{13X_3^2}{4}\bigg)z^2 +\bigg(-15(X_1^{(i_1)})^2 +39X_1^{(i_1)}X_2^{(i_2)}-14X_1^{(i_1)}X_3 -24(X_2^{(i_2)})^2+17X_2^{(i_2)}X_3-3X_3\bigg)z + \nonumber \\
& \bigg(9(X_1^{(i_1)})^2 -18X_1^{(i_1)}X_2^{(i_2)} +6X_1^{(i_1)}X_3+9(X_2^{(i_2)})^2-6X_2^{(i_2)}X_3+X_3^2\bigg)
\end{align}
\end{table*}
For each $[i_1,i_2]\in \{[1,1],[1,2],[2,1],[2,2]\}$, $u_0,\dots,u_4$ for this example are evident in \eqref{f o u in example 1}. 
\end{example}
\begin{definition}[monomial vector]
Let $\mathcal{D}(f)$ be the set that contains the degrees in $f$\footnote{For example, $\mathcal{D}(f)=\{0,1,3\}$ for $f(x)=x^3+x+1$.}. The \textit{monomial vector} that contains degree $d\in\mathcal{D}_f$ monomials of $X_k, k\in \mathcal{H}_{\mathcal{K}}$ and $X_k^{(\mathbf{v}[k])}, k\in\mathcal{A}_{\mathcal{K}}, \mathbf{v}\in[v]^{\beta'}$, is defined as 
\begin{align}
    \mathbf{X} \coloneqq \bigcup_{\substack{ \sum_{k\in[K]}i_k=d\in \mathcal{D}(f) \\  \mathbf{v}\in[v]^{\beta'}}}\bigg[\prod_{\substack{k\in\mathcal{A}_{\mathcal{K}}\\k'\in\mathcal{H}_{\mathcal{K}}}}(X_k^{(\mathbf{v}[k])})^{i_k}(X_{k'})^{i_{k'}}\bigg],
\end{align}
where $\cup$ indicates concatenation.
\end{definition}
For Example \ref{example 1},
\begin{align}\label{X in example 1}
    \mathbf{X} =& [(X_1^{(1)})^2,~ (X_1^{(2)})^2, ~ (X_2^{(1)})^2,~ (X_2^{(2)})^2, ~ (X_3)^2, ~ X_1^{(1)}X_2^{(1)}, \nonumber\\
    & ~ X_1^{(2)}X_2^{(1)}, X_1^{(1)}X_2^{(2)}, ~ X_1^{(2)}X_2^{(2)}, 
    X_1^{(1)}X_3, ~ X_1^{(2)}X_3, \nonumber\\
    & X_2^{(1)}X_3, ~ X_2^{(2)}X_3]
\end{align}
The length of $\mathbf{X}$ is
\begin{align}\label{L definition}
    L = \sum_{d\in\mathcal{D}(f)}\sum_{m=0}^{d}\binom{\beta'}{m}\binom{K-\beta'-1+d}{d-m}v^m
\end{align}
Each $u_i(X_r^{(\mathbf{v}[r])},X_s,~ r\in \mathcal{A}_{\mathcal{K}},s\in \mathcal{H}_{\mathcal{K}})$, $0\le i\le d(K-1), \mathbf{v}\in[v]^{\beta'}$ is a linear combination of the elements of $\mathbf{X}$. So we can define the \textit{characteristic matrix}.
\begin{definition}[characteristic matrix] \label{M definition}
The \textit{characteristic matrix} of Lagrange encoding, denoted by $\mathbf{M}_{\textrm{Lagrange}}$, is a matrix of dimension $\big(v^{\beta'}(d(K-1)+1)\big)\times L$, that illustrates the relation between the coefficient vector and the monomial vector, i.e.
\begin{align}
    \mathbf{U}_{\textrm{Lagrange}}=\mathbf{M}_{\textrm{Lagrange}}\mathbf{X}
\end{align}
\end{definition}
Note that even though we defined the coefficient vector and characteristic matrix for Lagrange encoding, any encoding in the form of \eqref{encoding polynomial q} has an associated coefficient vector and characteristic matrix.

For Example \ref{example 1}, $\mathbf{M}_{\textrm{Lagrange}}$ can be found by using \eqref{f o u in example 1} and \eqref{X in example 1}, and is shown in \eqref{M for example 1}. Generally, the characteristic matrix is only a function of $\mathcal{A}_{\mathcal{K}}$, $\mathcal{H}_{\mathcal{K}}$, $q^{(\mathbf{v})}$, i.e. encoding scheme, $v$, and $f$. In fact, $\mathbf{M}$ is not dependent on the adversarial behavior, because it simply contains the relationships between all of adversarial behaviors. Therefore, given the adversarial set, encoding polynomial, $v$, and $f$, we can find $\mathbf{M}$, and its rank. That is why we named it the characteristic matrix.
\begin{figure*}
    \centering
\begin{align}\label{M for example 1}
\left[\begin{array}{c}
     u_4(X_1^{(1)},X_2^{(1)},X_3) \\
     u_3(X_1^{(1)},X_2^{(1)},X_3) \\
     u_2(X_1^{(1)},X_2^{(1)},X_3) \\
     u_1(X_1^{(1)},X_2^{(1)},X_3) \\
     u_0(X_1^{(1)},X_2^{(1)},X_3) \\ \hline
     u_4(X_1^{(1)},X_2^{(2)},X_3) \\
     u_3(X_1^{(1)},X_2^{(2)},X_3) \\
     u_2(X_1^{(1)},X_2^{(2)},X_3) \\
     u_1(X_1^{(1)},X_2^{(2)},X_3) \\
     u_0(X_1^{(2)},X_2^{(2)},X_3) \\ \hline
     u_4(X_1^{(2)},X_2^{(1)},X_3) \\
     u_3(X_1^{(2)},X_2^{(1)},X_3) \\
     u_2(X_1^{(2)},X_2^{(1)},X_3) \\
     u_1(X_1^{(2)},X_2^{(1)},X_3) \\
     u_0(X_1^{(2)},X_2^{(1)},X_3) \\ \hline
     u_4(X_1^{(2)},X_2^{(2)},X_3) \\
     u_3(X_1^{(2)},X_2^{(2)},X_3) \\
     u_2(X_1^{(2)},X_2^{(2)},X_3) \\
     u_1(X_1^{(2)},X_2^{(2)},X_3) \\
     u_0(X_1^{(2)},X_2^{(2)},X_3)
\end{array}\right]= \left[\begin{array}{cc|cc|c|cccc|cc|cc}
     \frac{1}{4} & 0 & 1 & 0 & \frac{1}{4} & -1 & 0 & 0 & 0 & \frac{1}{2} & 0 & -1 & 0 \\
     -\frac{5}{2} & 0 & -8 & 0 & -\frac{3}{2} & 9 & 0 & 0 & 0 & -4 & 0 & 7 & 0 \\
     \frac{37}{4} & 0 & 22 & 0 & \frac{13}{4} & -29 & 0 & 0 & 0 & \frac{23}{2} & 0 & -17 & 0 \\
     -15 & 0 & -24 & 0 & -3 & 39 & 0 & 0 &  0 & -14 & 0 & -17 & 0 \\
     9 & 0 & 9 & 0 & 1 & -18 & 0 & 0 & 0 & 6 & 0 & -6 & 0 \\ \hline
     \frac{1}{4} & 0 & 0 & 1 & \frac{1}{4} & 0 & 0 & -1 & 0 & \frac{1}{2} & 0 & 0 & -1 \\
     -\frac{5}{2} & 0 & 0 & -8 & -\frac{3}{2} & 0 & 0 & 9 & 0 & -4 & 0 & 0 & 7 \\
     \frac{37}{4} & 0 & 0 & 22 & \frac{13}{4} & 0 & 0 & -29 & 0 & \frac{23}{2} & 0 & 0 & -17 \\
     -15 & 0 & 0 & -24 & -3 & 0 & 0 & 39 & 0 & -14 & 0 & 0 & 17 \\
     9 & 0 & 0 & 9 & 1 & 0 & 0 & -18 & 0 & 6 & 0 & 0 & -6 \\ \hline
     0 & \frac{1}{4} & 1 & 0 & \frac{1}{4} & 0 & -1 & 0 & 0 & 0 & \frac{1}{2} & -1 & 0 \\
     0 & -\frac{5}{2} & -8 & 0 & -\frac{3}{2} & 0 & 9 & 0 & 0 & 0 & -4 & 7 & 0 \\
     0 & \frac{37}{4} & 22 & 0 & \frac{13}{4} & 0 & -29 & 0 & 0 & 0 & \frac{23}{2} & -17 & 0 \\
     0 & -15 & -24 & 0 & -3 & 0 & 39 & 0 & 0 & 0 & -14 & 17 & 0 \\
     0 & 9 & 9 & 0 & 1 & 0 & -18 & 0 & 0 & 0 & 6 & -6 & 0 \\ \hline
     0 & \frac{1}{4} & 0 & 1 & \frac{1}{4} & 0 & 0 & 0 & -1 & 0 & \frac{1}{2} & 0 & -1 \\
     0 & -\frac{5}{2} & 0 & -8 & -\frac{3}{2} & 0 & 0 & 0 & 9 & 0 & -4 & 0 & 7 \\
     0 & \frac{37}{4} & 0 & 22 & \frac{13}{4} & 0 & 0 & 0 & -29 & 0 & \frac{23}{2} & 0 & -17 \\
     0 & -15 & 0 & -24 & -3 & 0 & 0 & 0 & 39 & 0 & -14 & 0 & 17 \\
     0 & 9 & 0 & 9 & 1 & 0 & 0 & 0 & -18 & 0 & 6 & 0 & -6
\end{array}\right]\left[\begin{array}{c}
     (X_1^{(1)})^2 \\ (X_1^{(2)})^2 \\ \hline
     (X_2^{(1)})^2 \\ (X_2^{(2)})^2 \\ \hline
     (X_3)^2 \\ \hline
     X_1^{(1)}X_2^{(1)} \\ X_1^{(2)}X_2^{(1)} \\ X_1^{(1)}X_2^{(2)} \\ X_1^{(2)}X_2^{(2)} \\ \hline X_1^{(1)}X_3 \\ X_1^{(2)}X_3 \\ \hline
     X_2^{(1)}X_3 \\ X_2^{(2)}X_3
\end{array}\right].
\end{align}
\end{figure*}

Now we present our main result.

\begin{theorem}\label{main result}
Consider a system of $N$ worker nodes, $\beta$ of which are adversaries, and $K$ of which are message producers. $\beta'$ of the message producers are adversarial, each able to diffuse $v$ different messages in the system. If $\textrm{rank}(\mathbf{M}_{\textrm{Lagrange}})\le v^{\beta'}d(K-1)+1$, the threshold for recovering $f$ of the messages of the honest producers using Lagrange coded computing and linear decoding is $N^* \ge \textrm{rank}(\mathbf{M}_{\textrm{Lagrange}})$,
where $\mathbf{M}_{\textrm{Lagrange}}$ is the characteristic matrix, and $f$ is a polynomial of degree $d$.
\end{theorem}
The definition of the recovery threshold implies that $N\ge N^*$ is necessary, otherwise, $f(X_k), k\in\mathcal{H}_{\mathcal{K}}$, cannot be decoded. 
\begin{remark}
This result is based on the assumption that adversarial nodes broadcast $y_n=f(\Tilde{X}_n)$, $n\in\mathcal{A}$, i.e. they do not alter the final result. Since we introduce a lower bound on $N^*$, this assumption is justified.
\end{remark}
\begin{remark}
Even though the above theorem is stated for Lagrange computing of $f$, the attack is valid for other methods of coded computing. However, the recovery threshold depends on the deployed coded computing method. In particular, if the encoding is in the form of \eqref{encoding polynomial q}, with different $L$ coefficients, and possibly different degree, the result of Theorem \ref{main result} holds by replacing the corresponding characteristic matrix with $\mathbf{M}_{\textrm{Lagrange}}$.
\end{remark}
\begin{remark}
In order to use this result for Polyshard, suppose that a shard is controlled by adversaries if $\gamma\in[0,1]$ fraction of its nodes are controlled by adversaries. In sharding, there are $N$ nodes and $K$ shards, so there are $\frac{N}{K}$ nodes in each shard. Consequently, adversaries can take over a maximum of $\beta'=\frac{\beta}{\gamma\frac{N}{K}}$ shards. The adversarial shards are equivalent to the adversarial message producers in the general model. If we set $\beta=cN$ for a constant $c\in\mathbb{N}$, we have $\beta'=\frac{K}{c\gamma}$. In order to achieve scalability, Polyshard sets $K=O(N)$, but this also sets $\beta'=O(N)$. Consequently, $L\approx O(K^d)$ according to \eqref{L definition}. Since $\textrm{rank}(\mathbf{M}_{\textrm{Lagrange}})\le L$, and $L\ll v^{\beta'}d(K-1)+1$, the condition in Theorem \ref{main result} is automatically satisfied, making its result applicable to this case. In Lemma \ref{rank lemma} that follows, we show that for the mentioned parameters of Polyshard, $\textrm{rank}(\mathbf{M}_{\textrm{Lagrange}})\ge O(K^{\frac{d}{2}})$. Thus $N^*\ge O(K^{\frac{d}{2}})$ according to
Theorem \ref{main result}, which means that Polyshard cannot be secure and linearly scale with $N$ at the same time.
\end{remark}
\begin{remark}
Setting $v=1$ means no effective adversary exists.
In this case, $\mathbf{\mathbf{M}}_{\textrm{Lagrange}}$ is of dimension $(d(K-1)+1) \times L$, and $d(K-1)+1\ll L$. It is easy to verify that if $\omega_k, k\in[K]$ are chosen independently and uniformly at random from $\mathbb{F}$, $\textrm{rank}(\mathbf{M}_{\textrm{Lagrange}}) = d(K-1)+1$. Therefore, the lower bound in Theorem \ref{main result} gives the exact recovery threshold of normal LCC, i.e. $d(K-1)+1$.
\end{remark}
\begin{remark}
The first summation in \eqref{q^V definition} that contains the messages of the adversaries, can have $v^{\beta'}$ different combinations, counting $v$ cases for each $X^{(\mathbf{v}[k])}_k, k\in\mathcal{A}_{\mathcal{K}}$. Thus, there are $v^{\beta'}$ different $q^{(\mathbf{v})}(z)$ polynomials. If $N\ge v^{\beta'}(d(K-1)+1)$, and we know which polynomial each node has evaluated, we can pick out at least $(d(K-1)+1)$ consistent evaluations of one polynomial using which we can decode $f(X_k), k\in\mathcal{H}_{\mathcal{K}}$. Therefore, $N^*\le v^{\beta'}(d(K-1)+1)$, given the mentioned knowledge of the adversarial behaviour.
\end{remark}
Theorem \ref{main result} statement uses the notion $\textrm{rank}(\mathbf{M}_{\textrm{Lagrange}})$. In the following lemma, we provide a lower bound on $\textrm{rank}(\mathbf{M}_{\textrm{Lagrange}})$ for the region of parameters that correspond to a sharded blockchain, as Polyshard.
\begin{lemma}\label{rank lemma}
Consider a system of $N$ worker nodes, $K=O(N)$ message producers, and $\beta=\beta'=O(K)$ adversarial message producer nodes, where $N$ is very large. In this regime, $\textrm{rank}(\mathbf{M}_{\textrm{Lagrange}})\ge O(k^{\frac{d}{2}})$, where $\mathbf{M}_{\textrm{Lagrange}}$ is the characteristic matrix introduced in Definition \ref{M definition}.
\end{lemma}
Proof of this lemma can be found in Appendix \ref{proof of rank lemma}.

\subsection{Proof of Theorem \ref{main result}}
We choose an arbitrary set $\mathcal{\hat{N}}\subseteq\mathcal{N}$ of size $\hat{N} = \textrm{rank}(\mathbf{M}_{\textrm{Lagrange}})-1$, and show that under some adversarial behaviours, $f(X_k), k\in\mathcal{H}_{\mathcal{K}}$ cannot be decoded from $y_n, n\in\hat{N}$, even if we know the adversarial behaviour, i.e. what each node have received from each adversary. In other words, we assume that we know the adversarial behaviour, and want to decode $f(X_k), k\in\mathcal{H}_{\mathcal{K}}$. Note that in general, we do not have such information, so the minimum number of equations needed for decoding under this assumption (which we show is greater than $\hat{N}$) is in fact a lower bound on the recovery threshold, i.e. $N^*>\hat{N}$. In the following, we parse the proof into some steps.

\textbf{Step $1$}. For any adversarial behavior, we can partition $\mathcal{\hat{N}}=\mathcal{N}_1\cup\dots\cup\mathcal{N}_{v^{\beta'}}$, where $n\in\mathcal{N}_i, i\in[v^{\beta'}]$ denotes the set of nodes that all receive the same versions of messages, $\mathbf{v}_i\in[v]^{\beta'}$, from $\mathcal{A}_{\mathcal{K}}$. Thus, 
\begin{align}
    f(q^{(\mathbf{v}_i)}(\alpha_n))&=y_n, \quad n\in\mathcal{N}_i, i\in[v^{\beta'}]. \label{y equations}
\end{align}
The values of $N_i\coloneqq|\mathcal{N}_i|, i\in[v^{\beta'}]$, show the adversarial behaviour. We are going to find $N_1,\dots,N_{v^{\beta'}}$ such that $N_1+\dots+N_{v^{\beta'}}=\textrm{rank}(\mathbf{M}_{\textrm{Lagrange}})-1$, but $f(X_k), k\in\mathcal{H}_{\mathcal{K}}$ cannot be decoded. 

\textbf{Step $2$}. Let us define $\textrm{Van}_{\mathcal{S}}^D$ as a $|\mathcal{S}|\times(D+1)$ Vandermonde matrix whose rows consist of elements of the set $\mathcal{S}$, from degree $0$ to $D$, for a $D\in\mathbb{N}$. For example, 
\begin{align*}
    \textrm{Van}_{\{\alpha_1,\alpha_2\}}^3=\begin{bmatrix}
    \alpha_1^3 & \alpha_1^2 & \alpha_1 & 1 \\
    \alpha_2^3 & \alpha_2^2 & \alpha_2 & 1
    \end{bmatrix}.
\end{align*}
We also define $y_{\mathcal{S}}$ as a vector that consists of the elements $y_n, n\in \mathcal{S}$, for the set $\mathcal{S}$. The matrix equivalence of \eqref{y equations} is
\begin{align}\label{y equations matrix form}
\mathbf{A}\mathbf{U}=\mathbf{y}\coloneqq\begin{bmatrix}
y_{\mathcal{N}_1} \\
y_{\mathcal{N}_2} \\
\vdots \\
y_{\mathcal{N}_{v^{\beta'}}}
\end{bmatrix},
\end{align}
where 
\begin{align}\label{A definition}
\mathbf{A}\coloneqq\left[\begin{array}{ccc}
\textrm{Van}_{\{\alpha_n,n\in\mathcal{N}_1\}}^{d(K-1)} & \dots & 0 \\
& \vdots & \\
0 & \dots & \textrm{Van}_{\{\alpha_n,n\in\mathcal{N}_{v^{\beta'}}\}}^{d(K-1)}
\end{array}\right].
\end{align}
The dimension of $\mathbf{A}$ is $\hat{N}\times \bigg(v^{\beta'}(d(K-1)+1)\bigg)$. We define a vector $\mathbf{Z}\coloneqq[Z_k, k\in\mathcal{H}_{\mathcal{K}}]$, where $Z_k \coloneqq f(X_k), k\in\mathcal{H}_{\mathcal{K}}$. The length of $\mathbf{Z}$ is $k-\beta'$. The equations $Z_k \coloneqq f(X_k), k\in\mathcal{H}_{\mathcal{K}}$, can be expressed in the matrix form
\begin{align}\label{z equation matrix form}
\left[\begin{array}{c|c|c}
\textrm{Van}_{\{\omega_k,k\in\mathcal{H}_{\mathcal{K}}\}}^{d(K-1)}   & \mathbf{0}_{(K-\beta')\times(v^{\beta'}-1)(d(K-1)+1)}       &  -\mathbf{I}_{K-\beta'}
    \end{array}\right]
    \begin{bmatrix}
     \mathbf{U}\\
     \mathbf{Z}
    \end{bmatrix} = \mathbf{0}.
\end{align}
For the above equations, we used the fact that $f(X_k)=f(q^{\mathbf{v}_1}(\omega_k)), k\in\mathcal{H}_{\mathcal{K}}$, therefore the first 
$d(K-1)+1$ variables in $\mathbf{U}$ are present in \eqref{z equation matrix form}. Note that we could have used $f(X_k)=f(q^{\mathbf{v}_i}(\omega_k)), k\in\mathcal{H}_{\mathcal{K}}$, for any $i\in[v^{\beta'}]$, because 
\begin{align}
f(q^{(\mathbf{v}_1)}(\omega_k)) =  \dots=f(q^{(\mathbf{v}_{v^{\beta'}})}(\omega_k)), \quad k\in\mathcal{H}_{\mathcal{K}}.
\end{align}
\textbf{Step $3$}. The next step is finding the linear dependencies between the elements of $\mathbf{U}_{\textrm{Lagrange}}$. We know that all the $v^{\beta'}$ polynomials,  $f(q^{(\mathbf{v}_1)}(z)), \dots,f(q^{(\mathbf{v}_{v^{\beta'}})}(z))$, have the same value at each point $\omega_k, k\in\mathcal{H}_{\mathcal{K}}$. This give us the hint that these polynomials, and thus their coefficients which are in $\mathbf{U}$,
are related to each other. Thus, We want to find a full rank matrix $\mathbf{P}_{\textrm{Lagrange}}$ such that $\mathbf{P}_{\textrm{Lagrange}}\mathbf{U}_{\textrm{Lagrange}}=\mathbf{0}$, for all values of $X_k^{(i)}\in\mathbb{F}, k\in\mathcal{A}_{\mathcal{K}},i\in[v]$, and $X_k\in\mathbb{F}, k\in\mathcal{H}_{\mathcal{K}}$. Since $\mathbf{U}_{\textrm{Lagrange}}=\mathbf{M}\mathbf{X}$, $\mathbf{P}_{\textrm{Lagrange}}\mathbf{U}_{\textrm{Lagrange}}=\mathbf{0}$ is equivalent to $\mathbf{P}_{\textrm{Lagrange}}\mathbf{M}_{\textrm{Lagrange}}=\mathbf{0}$. As a result, $\mathbf{P}$ is made of the basis of the left null space of $\mathbf{M}_{\textrm{Lagrange}}$. The dimension of $\mathbf{P}_{\textrm{Lagrange}}$ is
\begin{align*}
    \big(v^{\beta'}(d(K-1)+1)-\textrm{rank}(\mathbf{M}_{\textrm{Lagrange}})\big) \times \big(v^{\beta'}(d(K-1)+1)\big).
\end{align*}
For Example \ref{example 1}, $\mathbf{M}_{\textrm{Lagrange}}$ is a $20\times 13$ matrix and $\textrm{rank}(\mathbf{M}_{\textrm{Lagrange}})=12$. So $\mathbf{P}_{\textrm{Lagrange}}$ is a $8\times 20$ full row rank matrix.  \\
\textbf{Step $4$}. All the equations we have are summarized in
\begin{align}\label{all equations}
\mathbf{D}\begin{bmatrix}
     \mathbf{U}_{\textrm{Lagrange}}\\
     \mathbf{Z}
    \end{bmatrix} = \begin{bmatrix}
     \mathbf{y}\\
     \mathbf{0}
    \end{bmatrix}.
\end{align}
where 
\begin{align*}
\mathbf{D}\coloneqq\left[\begin{array}{c|c|c}
\multicolumn{2}{c|}{\mathbf{A}}          & \mathbf{0}          \\ \hline
\multicolumn{2}{c|}{\mathbf{P}}          & \mathbf{0}          \\ \hline
\textrm{Van}_{\{\omega_k,k\in\mathcal{H}_{\mathcal{K}}\}}^{d(K-1)}   & \mathbf{0}       &  -\mathbf{I}_{K-\beta'}
    \end{array}\right],
\end{align*}
and
\begin{align*}
\mathbf{D}_1\coloneqq\left[\begin{array}{c|c}
\multicolumn{2}{c}{\mathbf{A}}            \\ \hline
\multicolumn{2}{c}{\mathbf{P}}         \\ \hline
\textrm{Van}_{\{\omega_k,k\in\mathcal{H}_{\mathcal{K}}\}}^{d(K-1)}   & \mathbf{0}  
    \end{array}\right], \mathbf{D}_2\coloneqq\left[\begin{array}{c}
 \mathbf{0}          \\ \hline
 \mathbf{0}          \\ \hline
 -\mathbf{I}_{K-\beta'}
    \end{array}\right]
\end{align*}
The condition for the unique determination of $\mathbf{Z}$ in \eqref{all equations} is $\textrm{rank}(\mathbf{D})=\textrm{rank}(\mathbf{D}_1)+\textrm{rank}(\mathbf{D}_2)$. This ensures that $\mathbf{D}_2$ does not lie in the column space of $\mathbf{D}_1$, and \eqref{all equations} can be converted to a full rank set of equations of only $\mathbf{Z}$, where $\mathbf{Z}$ can be found uniquely. \\
\textbf{Step $5$}. Recall that we want to find $N_1,\dots,N_{v^{\beta'}}$ such that $N_1+\dots+N_{v^{\beta'}}=\textrm{rank}(\mathbf{M}_{\textrm{Lagrange}})-1$, and $\mathbf{Z}$ cannot be determined uniquely.
We choose the adversarial behaviour where $N_1,\dots,N_{v^{\beta'}}\le d(K-1)$, and $N_1+\dots+N_{v^{\beta'}}=\textrm{rank}(\mathbf{M}_{\textrm{Lagrange}})-1$. This is why the condition $\textrm{rank}(\mathbf{M}_{\textrm{Lagrange}})\le v^{\beta'}d(K-1)+1$, that was stated in the theorem is needed.

With this choice of $N_1,\dots,N_{v^{\beta'}}$, the Vandermonde matrices in $\mathbf{F}$, 
$\textrm{Van}_{\{\alpha_n,n\in\mathcal{N}_1\}}^{d(K-1)},\dots,\textrm{Van}_{\{\alpha_n,n\in\mathcal{N}_{v^{\beta'}}\}}^{d(K-1)}$ are full row rank, and  consequently, $\mathbf{A}$ is full row rank as well. But the important point is that $\mathbf{A}\mathbf{U}=\mathbf{y}$ is an underdetermined system, so $\mathbf{Z}$ cannot be easily deducted from \eqref{z equation matrix form}.

The matrix $\mathbf{A}$ is exclusively formed with $\alpha_1,\dots,\alpha_N$, and $\mathbf{P}$ is exclusively formed with $\omega_1,\dots,\omega_K$. We know that $\alpha_1,\dots,\alpha_N, \omega_1,\dots,\omega_K$ are all chosen independently and uniformly from $\mathbb{F}$. Moreover, the dimension of $\mathbf{A}$ is 
\begin{align*}
    \bigg(\textrm{rank}(\mathbf{M}_{\textrm{Lagrange}})-1\bigg)\times \bigg(v^{\beta'}(d(K-1)+1)\bigg),
\end{align*}
and the dimension of $\mathbf{P}$ is
\begin{align*}
    \bigg(v^{\beta'}(d(K-1)+1)-\textrm{rank}(\mathbf{M}_{\textrm{Lagrange}})\bigg) \times \bigg(v^{\beta'}(d(K-1)+1)\bigg).
\end{align*}
Therefore, $[\mathbf{A}^T|\mathbf{P}^T]^T$ is full row rank with dimension $\big(v^{\beta'}(d(K-1)+1)-1\big)\times \big(v^{\beta'}(d(K-1)+1)\big)$. 

Consider an arbitrary row of $[\textrm{Van}_{\{\omega_k,k\in\mathcal{H}_{\mathcal{K}}\}}^{d(K-1)}|\mathbf{0}]$, say $\mathbf{r}=[(\omega_k)^{d(K-1)} \dots \omega_k~1 | \mathbf{0}]$ for a $k\in\mathcal{H}_{\mathcal{K}}$, and let its index be $m\in[K-\beta']$.
This row is not in the row span of $\mathbf{P}$, because $\mathbf{r}\mathbf{M}_{\textrm{Lagrange}}\neq \mathbf{0}$. It is not in the row span of $\mathbf{A}$ either, because 
$N_1\le d(K-1)$, and $\textrm{Van}_{\{\alpha_n,n\in\mathcal{N}_1\}}^{d(K-1)}$ is underdetermined. Therefore, $[\mathbf{A}^T|\mathbf{P}^T|\mathbf{r}^T]^T$ is a full rank square matrix of dimension $v^{\beta'}(d(K-1)+1)$. As a result, the vector $[0~ \dots~ 0~ 1]$ -all zeros except the last element- of dimension $v^{\beta'}(d(K-1)+1)$, is in the column span of $[\mathbf{A}^T|\mathbf{P}^T|\mathbf{r}^T]^T$. This violates $\textrm{rank}(\mathbf{D})=\textrm{rank}(\mathbf{D}_1)+\textrm{rank}(\mathbf{D}_2)$, because it means that $m$'th column of $\mathbf{D}_2$ is in the column span of $\mathbf{D}_1$. Hence, $\mathbf{Z}$ cannot be found uniquely, and the proof of Theorem \ref{main result} is complete.

It is worth noting that if there were a mechanism that let us know the adversarial behavior when decoding, i.e. we knew what each node have received from each adversarial node, the bound given in Theorem \ref{main result} would be tight, and $\textrm{rank}(\mathbf{M}_{\textrm{Lagrange}})$ equations would be enough for recovering $f(X_k), k\in\mathcal{H}_{\mathcal{K}}$.

\section{A partial solution to the discrepancy attack}
In this section, we propose a partial solution that help us detect the adversaries and then decode $f(X_k), k\in\mathcal{H}_{\mathcal{K}}$ reliably. This solution modifies Polyshard minimally, and preserves its communication load of $O(N^2)$.

As mentioned before, if unlike reality, the communication load was not an issue, the discrepancy attack could be easily detected by forcing nodes to broadcast their received blocks. Since valid blocks have signatures in them, no adversary can deny the fact that it has produced and sent more than one block to nodes. This vanilla method has a total load of $O(KN^2)$, or equivalently $O(N^3)$ if we set $K=O(N^2)$, so we need better detection schemes in terms of communication load.

In the following, we explain our proposed solution, which we call the \textit{modified Polyshard} hereafter. 
Let $X_{k,n}$ be the block of shard $k\in[K]$ that node $n\in[N]$ receives. We add a new round, after shards broadcast their blocks, and before any computation, in which each node $n$ chooses one of $X_{1,n},\dots,X_{K,n}$, uniformly at random, named $\dot{X}_n$, and broadcasts it.
At the end of this round, one of these cases happen:
\begin{itemize}
    \item 
    There are two honest nodes, $n_1,n_2\in[N]$, whose chosen blocks, $\dot{X}_{n_1}$ and $\dot{X}_{n_2}$, are different but belong to the same shard $k\in[K]$. Therefore, all nodes are alerted that shard $k$ is adversarial. So all nodes discard $X_{k,n}$ from all subsequent steps.
    \item
    Shard $k\in[K]$ is adversarial, but there is only one honest node $n\in[N]$ that chooses $\dot{X}_n=X_{k,n}$. In this case, nodes that had received a something other than $\dot{X}_n$ from shard $k$ in the first round, realize the attack, whereas nodes that had received the same block remain clueless. Since the goal is to get rid of the discrepancy, if at the end of this round, a node $n'\in[N]$ finds a single $\dot{X}_n$ for shard $k$ such that $\dot{X}_n\neq X_{k,n'}$, it should use $\dot{X}_n$ instead of $X_{k,n'}$ in the subsequent steps. In this way, nodes that had received different versions of a block use the same block as other nodes in the network, and the discrepancy is resolved.
    \item
    Shard $k\in[K]$ is adversarial, but no honest node chooses $k$'th received block. Thus, the discrepancy cause by this shard remains undetected.
\end{itemize}
The above procedure resolves the discrepancies cause by the adversaries, if for each adversarial shard, there is at least one honest node that chooses that shard. This is because either all honest nodes detect an adversarial shard, or, some shards detect an adversarial shard and align themselves with those that did not.

Even though the new proposed round adds to the total execution time, since the size of a block is constant, the communication load remains $O(N^2)$. Moreover, the adversaries cannot broadcast incorrect and misleading information, because of the signatures in blocks. 

In the following theorem, we set some values for $N$, $K$, and $\beta$, that ensures all the discrepancies caused by the adversarial shards are resolved with high probability. 
\begin{theorem}
Consider a blockchain system based on Polyshard with $N=K\ln K\in\mathbb{N}$ nodes, $K\in\mathbb{N}$ shards, $s\coloneqq\ln K\in\mathbb{N}$ nodes in a shard, and $\beta=c K\in\mathbb{N}$ adversarial nodes. We consider a shard adversarial if $\gamma\in[0,1]$ fraction of its nodes are adversarial. The probability that all the discrepancies cause by the adversarial shards are resolved in modified Polyshard goes to $1$ as $K,N\rightarrow \infty$.
\end{theorem}
The values chosen in this theorem are different from the values in \cite{li2020polyshard}. Indeed, we have decreased $K$ and $\beta$ from $O(N)$ in \cite{li2020polyshard}, to $O(e^{\mathbf{W}(N)})$, to be able to resolve the discrepancies, where $\mathbf{W}(.)$ is Lambert W function\footnote{The definition of Lambert W function is $y=W(x)$ if $ye^y=x$. We can write $N=K\ln K$ as $N=e^{\ln K}\ln K$, so $\ln K=\mathbf{N}(N)$, or $K=e^{\mathbf{W}(N)}$}. For $1\ll N$, $\mathbf{W}(N)\ll N$. While this means less scalability and less security, we need some sort of compromise to detect the discrepancy attack and prevent Polyshard from falling apart. 
\begin{proof}
The maximum number of adversarial shards is
\begin{align}\label{beta'}
    \beta' = \frac{\beta}{\gamma s} = \frac{c}{\gamma}\frac{K}{\ln K}.
\end{align}
Without loss of generality, assume that the first $\beta'$ shards of $K$ shards are adversarial. Each honest node $n\in[N]$ chooses an index $J_n\in[K]$ uniformly at random, and independently from the other nodes, and sets $\dot{X}_n = X_{J,n}$. We want to find the probability that each element of $[\beta']$ is chosen by at least one honest node. This is the famous \textit{coupon collector's problem}, and we use the standard solution for it here.

Let $G_i$, $i\in[\beta']$ be the random variable that indicates the number of honest nodes needed until the last node chooses $i$'th new adversarial shard. Then, $G=G_1+\dots+G_{\beta'}$ is a random variable that indicates the number of required honest nodes such that each element of $[\beta']$ is picked at least once. Because we set $N\rightarrow \infty$, we know that $G$ is concentrated around $\mathbb{E}(G)$ with a probability approaching $1$. The number of honest nodes should be greater than $\mathbb{E}(G)$. There are $N-\beta = N-cK$ honest nodes.

$G_1,\dots,G_{\beta'}$ are geometric random variables with $p_1=\frac{\beta'}{K}$, $p_2=\frac{\beta'-1}{K}, \dots, p_{\beta'}=\frac{1}{K}$. We know that $\mathbb{E}(G_i)=\frac{1}{p_i}=\frac{K}{\beta'+1-i}$. Therefore 
\begin{align}
    \mathbb{E}(G) = K(\frac{1}{\beta'}+\dots+\frac{1}{2}+1) \approx K \ln \beta'
\end{align}
Using \eqref{beta'},
\begin{align}
    \mathbb{E}(G) &= K \ln (\frac{c}{\gamma}\frac{K}{\ln K})= K\ln K + K \ln \frac{c}{\gamma} - K\ln\ln K \nonumber\\
    & = N + K \ln \frac{c}{\gamma} - K\ln\ln K
\end{align}
Therefore, the condition $\mathbf{E}(G)\le N-cK$ is equivalent to
\begin{align*}
    \ln\frac{c}{\gamma}-\ln\ln K\le -c,
\end{align*}
or
\begin{align*}
    c+ \ln\frac{c}{\gamma} \le \ln\ln K,
\end{align*}
which easily holds as $K\rightarrow\infty$. Therefore, with high probability, there are enough honest nodes that choose the adversarial shards in the new second step, and resolve the discrepancies.
\end{proof}

\begin{appendices}
\section{Proof of Lemma \ref{rank lemma}}\label{proof of rank lemma}
In this proof, without loss of generality, we assume that $\mathcal{A}_{\mathcal{K}}=\{1,\dots,\beta\}$. We first prove the lemme for $f(X)=X^d$, and then generalize the proof for any degree $d$ polynomial.

First, we need to examine the characteristic matrix. This matrix is composed of $v^{\beta}$ \textit{blocks}, and each block is a $(d(K-1)+1)\times L$ submatrix associated with a $\beta$-tuple, $\mathbf{v}=(v_1,\dots,v_{\beta})\in[v]^{\beta}$, indicating the versions of the messages of the adversaries. In fact, the block associated with $\mathbf{v}$ contains the coefficients of $f(q^{(\mathbf{v})}(z))$.

Since $\mathbf{U}=\mathbf{M}_{\textrm{Lagrange}}\mathbf{X}$, each column of $\mathbf{M}_{\textrm{Lagrange}}$ is associated with a monomial $(X_1^{(i_1)})^{d_1}\dots (X_{\beta}^{(i_{\beta})})^{d_{\beta}}(X_{\beta+1})^{d_{\beta+1}}\dots (X_K)^{d_K}$, where $d_1+\dots+d_K=d$, and $(i_1,\dots,i_{\beta})\in[v]^{\beta}$. In a block associated with $\mathbf{v}$, the subcolumns associated with monomials that \textit{contrast} $\mathbf{v}$ are zeros. By contrast, we mean that there is at least one $k\in[\beta]$ such that $d_k\neq 0$ and $i_k\neq v_k$, i.e. the versions of the adversarial messages in the monomial is not consistent with $\mathbf{v}$. In other words, a block associated with $\mathbf{v}$ has nonzero elements in columns corresponding to monomials that are consistent with $\mathbf{v}$.

In order to prove $\textrm{rank}(\mathbf{M}_{\textrm{Lagrange}})$ is greater than $\mathcal{O}(K^{\frac{d}{2}})$, it suffices to find a full rank submatrix of dimension $\mathcal{O}(K^{\frac{d}{2}})$ in $\mathbf{M}_{\textrm{Lagrange}}$. In the following, we carefully pick $\mathcal{O}(K^{\frac{d}{2}})$ columns and rows from $\mathbf{M}_{\textrm{Lagrange}}$ that form such full rank matrix.
Let 
\begin{align}\label{S1 definition}
\mathcal{S}_1\coloneqq \{X_{k_1}^{(1)}\dots X_{k_{\frac{d}{2}}}^{(1)},~~ (k_1,\dots,k_{\frac{d}{2}})\in \mathcal{R}^{[\beta]}_{\frac{d}{2}}\},
\end{align}
where $\mathcal{R}^{[\beta]}_{\frac{d}{2}}$ is the set of all $\frac{d}{2}$-combinations of $[\beta]$. So $|\mathcal{S}_1|=\binom{\beta}{\frac{d}{2}}$. Since $\beta \gg d$, $|\mathcal{S}_1|\approx \mathcal{O}(K^{\frac{d}{2}})$. We also define another set
\begin{align}\label{S2 definition}
\mathcal{S}_2\coloneqq \{(X_{\beta+1})^{d_{\beta+1}}\dots (X_K)^{d_K},~ d_{\beta+1}+\dots+d_K=\frac{d}{2}\}.
\end{align}
So $|\mathcal{S}_2|=\binom{K-\beta-1+\frac{d}{2}}{\frac{d}{2}}\approx \mathcal{O}(K^{\frac{d}{2}})$. 

We choose a submatrix in the following way. We choose $\mathcal{O}(K^{\frac{d}{2}})$ columns of $\mathbf{M}_{\textrm{Lagrange}}$ by choosing monomials that are product of one element from $\mathcal{S}_1$ and one element from $\mathcal{S}_2$. In other words, we choose the columns associated with the monomials in
\begin{align*}
\mathcal{S}\coloneqq& \{X_{k_1}^{(1)}\dots X_{k_{\frac{d}{2}}}^{(1)}(X_{\beta+1})^{d_{\beta+1}}\dots (X_K)^{d_K}, \\
& X_{k_1}^{(1)}\dots X_{k_{\frac{d}{2}}}^{(1)}\in \mathcal{S}_1, (X_{\beta+1})^{d_{\beta+1}}\dots (X_K)^{d_K}\in\mathcal{S}_2\},
\end{align*}
and use each element of $\mathcal{S}_1$ and $\mathcal{S}_2$ only once in $\mathcal{S}$. We choose blocks that are associated with $\beta$-tuples in
\begin{align*}
\mathcal{P} =& \{(v_1,\dots,v_{\beta}),~ v_i=1 ~\textrm{for}~ i\in r,\\
& v_i=2 ~\textrm{for}~ i\notin r ,~ r\in\mathcal{R}^{[\beta]}_{\frac{d}{2}+1})\},
\end{align*}
where $\mathcal{R}^{[\beta]}_{\frac{d}{2}+1}$ is the set of all $(\frac{d}{2}+1)$-combinations of $[\beta]$. Thus, $|\mathcal{P}|=\binom{\beta}{\frac{d}{2}+1}\approx \mathcal{O}(K^{\frac{d}{2}+1})$. In a block associated with $\mathbf{v}\in\mathcal{P}$, the subcolumns in $\mathcal{S}$ that are consistent with $\mathbf{v}$ are nonzero. There are $\frac{d}{2}$ subcolumns in $\mathcal{S}$ consistent with each $\mathbf{v}\in\mathcal{P}$. Those $\frac{d}{2}$ nonzero subcolumns in each block are linearly independent, assuming $\omega_1,\dots,\omega_K$ are chosen independently and uniformly at random from $\mathbb{F}$ in the Lagrange polynomial. Since the nonzero subcolumns in all blocks are linearly independent, the whole submatrix of blocks in $\mathcal{P}$ and columns in $\mathcal{S}$ are linearly independent. This submatrix is of dimension $\mathcal{O}(K^{\frac{d}{2}+1})\times \mathcal{O}(K^{\frac{d}{2}})$, so $\textrm{rank}(\mathbf{M}_{\textrm{Lagrange}})\ge \mathcal{O}(K^{\frac{d}{2}})$. This completes the proof for $f=X^d$.

For a general degree $d$ function, the monomial vector contains monomials of degrees in $\mathcal{D}_f$, which includes $d$. Thus, the submatrix described above still exists in the characteristic matrix, and proves that $\textrm{rank}(\mathbf{M}_{\textrm{Lagrange}})\ge \mathcal{O}(K^{\frac{d}{2}})$.

\end{appendices}

\bibliographystyle{IEEEtran}
\bibliography{ref}
\end{document}